\documentclass[preprint,12pt]{elsarticle}

\usepackage{amssymb}
\usepackage{color}

\journal{Physics Letters B}

\begin{document}

\begin{frontmatter}



\title{A First Experimental Limit on In-matter Torsion from Neutron Spin Rotation in Liquid ${}^4$He}


\author[IUCSS]{Ralf Lehnert}
\author[IUCSS,IU,CEEM]{W.M.~Snow}
\author[IU,CEEM]{H.~Yan}

\address[IUCSS]{Indiana University Center for Spacetime Symmetries, Bloomington, IN 47405, USA}
\address[IU]{Indiana University, Bloomington, IN 47405, USA}
\address[CEEM]{Center for Exploration of Energy and Matter,
Indiana University,\\
Bloomington, IN 47408, USA}

\begin{abstract}
We report the first experimental upper bound to our knowledge on possible in-matter torsion interactions of the neutron from a recent search for parity violation in neutron spin rotation in liquid $^{4}$He. Our experiment constrains a coefficient $\zeta$ consisting of a linear combination of parameters involving the time components of the torsion fields $T^\mu$ and $A^\mu$ from the nucleons and electrons in helium which violates parity. We report an upper bound of 
$|\zeta| < 9.1 \times 10^{-23}$ 
GeV at 68\% confidence level and indicate other physical processes that could be analyzed to constrain in-matter torsion.
\end{abstract}


\end{frontmatter}

\section{Introduction}

Einstein's theory of general relativity (GR) posits an intimate connection between the geometry of spacetime and its matter content: the presence of matter curves spacetime and conversely the motion of matter is affected by curvature. 
The success of GR has encouraged physicists to consider the geometric structure of spacetime as a legitimate object of physical inquiry. 
This idea
naturally raises the question regarding 
the significance and measurement of torsion---a 
second mathematical quantity besides curvature that 
characterizes geometries
and therefore further quantifies the interaction between 
matter and geometry.

In GR, gravity is interpreted as spacetime curvature and test-particle trajectories are geodesics. This elegant concept provides the present-day basis for our understanding of the classical gravitational field. Spacetime torsion, another natural geometric quantity which characterizes spacetime geometry, vanishes in GR. However, many models that extend GR include  nonvanishing torsion that is typically sourced by some form of spin density~\cite{torsion}. 
In such models, 
the natural coupling strength of torsion to spin 
is the same as that of curvature to energy--momentum. 
Whereas energy--momentum densities capable of producing appreciable curvature can clearly be identified and observed in nature, spin-density sources strong enough to generate measurable torsion effects are difficult 
to find or fabricate. 
The unknown range of torsion, 
which is effectively zero in some models, 
poses an additional obstacle for measurements.
Torsion effects are therefore judged to be extremely small 
and difficult to observe,
which has long discouraged experimental investigations in this field~\cite{torsion}.

We follow the empirical view pursued by others~\cite{Overview}
and simply treat the question of the presence of torsion as an issue to be answered by experiment. Many recent experiments in this field motivated by searches for possible new short-range interactions between nonrelativistic quantum particles~\cite{micro} have set limits on specific torsion theories. In addition, 
astrophysical observations~\cite{astro},  kaon interferometry~\cite{Kaons},
LHC data~\cite{LHC}, gravitational-wave detectors~\cite{GravWave}, and satellite-based gravity tests~\cite{GProbeB} have been analyzed to constrain torsion in specific models. Model-independent experimental bounds on  torsion also exist. Tight model-independent constraints on the size of long-range torsion fields have recently been set through the appropriate reinterpretation of experiments designed to search for Lorentz and CPT violation~\cite{krt08}. These studies searched for torsion fields generated by the spin density of some macroscopic object with the torsion source and the torsion probe separated by macroscopic distances~\cite{PrefFrTests,SunSource}. Torsion is treated here as an external field outside of experimental control
which selects preferred directions for local physics. 
The effective violation of Lorentz symmetry can then be used to search for torsion in localized experiments, for example through boosts or rotation of the torsion probe relative to the fixed torsion background~\cite{CL97,krt08,DataTables}.

In this work we provide to our knowledge the first experimental upper bound on what we call ``in-matter'' torsion in which the spatial separation of torsion source and probe is eliminated. This constrains a qualitatively different class of models with short-ranged torsion and therefore complements the stringent bounds from tests based on Lorentz and CPT violation. 
Polarized slow neutrons are known to be an excellent choice for such an experimental investigation~\cite{neutrons}. They constitute a massive polarized probe that can penetrate macroscopic amounts of matter due to their lack of electric charge and lack of ionizing interactions with matter, and they can also be used to perform sensitive polarization measurements using various types of interferometric methods. Our experiment employed transversely polarized slow neutrons that traversed a meter of liquid $^{4}$He. We constrain possible internal torsion fields of arbitrary range generated by the spin-$\textstyle{\frac{1}{2}}$ protons, neutrons, and electrons in the $^{4}$He that violate parity and hence cause the neutron spin to rotate transverse to its momentum. 

%The helium target possesses no net spin density, and one might naively expect that this implies that no torsion effects can be produced. We will show below that this is false. Just as no \lq\lq net chirality\rq\rq of a medium is required to see the optical activity of light in a chiral medium, likewise there is no requirement that the electrons or the nuclei in the ensemble of helium atoms in the liquid target be polarized in order for the polarized neutrons traversing the helium to be able to see torsion effects. Since a large fraction of possible torsion fields violate mirror symmetry and lead to a term in the slow-neutron optical potential proportional to $\vec{S} \cdot \vec{p}$, our constraint applies to a large fraction of possible torsion fields and therefore constitutes in our opinion a relatively model-independent constraint of general interest. 

The distance scales probed by the spin-rotation observable, 
which involves the neutron forward scattering amplitude from the medium, 
are not in principle limited by the neutron size~\cite{KarlTadic}. 
To see this consider the simple case of a nonrelativistic potential of the form 
\begin{equation}\label{eqn1}
V(r)=\frac{g^{2}}{2\pi}\frac{\exp{(-r/\lambda)}}{r}\vec{\sigma}\cdot\vec{v}\,, 
\end{equation}
where $\lambda$ is the interaction range, 
$\vec{S}=\vec{\sigma}/2$ is the spin of the polarized particle 
(in our case the neutron), 
$r$ is the distance between the two interacting particles, 
and $g$ is the coupling strength. 
As observed above, 
the $\vec{\sigma}\cdot\vec{v}$ factor violates parity 
and therefore causes a rotation of the plane of polarization 
of a transversely polarized slow neutron beam about its momentum 
as it moves through matter, 
as observed experimentally 
in the case of the neutron--nucleus weak interaction~\cite{Mic64, Forte80, Heckel82, Heckel84}. 
The rotation angle per unit length $d\phi_{PV}/dL$ 
of a neutron of wave vector $\vec{k}_{n}$ 
in a medium of number density $\rho$ is $d\phi_{PV}/dL=4\pi \rho f_{PV}/k_{n}$, 
where $f_{PV}$ is the forward limit of the parity-odd p-wave scattering amplitude. 
This relationship holds despite the fact that 
the range of the nucleon--nucleon weak interaction is smaller than 
the size of the neutron. 
Because $f_{PV}$ is proportional to the parity-odd correlation $\vec{\sigma}\cdot \vec{k}_{n}$, 
$d\phi_{PV}/dL$ is constant as $k_{n} \to 0$ in the absence of resonances~\cite{Sto82}. 
In this case one can apply the Born approximation 
to derive the relation between $f_{PV}$ and the parameters of the potential, and the spin rotation angle per unit length can be expressed directly in terms of the coupling, the range of the interaction, and the number density as follows: ${d\phi_{PV}/dL}=4g^{2}\rho\lambda^{2}$. 
This observable therefore is independent of the neutron's wavelength and constrains a product of the strength and range of the parity-odd interaction.

In our theoretical analysis we follow the model-independent approach to torsion interactions taken in Ref.~\cite{krt08}. In particular, 
we can neglect pure GR effects 
due to the lack of parity-violating spin interactions
and work in a flat spacetime background. Note that contrary to 
superficial expectations this limit also does {\em not} imply vanishing torsion. We assume that inside the liquid $^{4}$He a torsion field ${\cal T}^{\alpha}{}_{\mu\nu}(x)$ is generated by the ambient spin density of the helium atoms. The detailed form of ${\cal T}^{\alpha}{}_{\mu\nu}(x)$ is model dependent, but it is reasonable to approximate the dominant effects by a spacetime-constant torsion background 
$\big\langle\, {\cal T}^{\alpha}{}_{\mu\nu}(x)\,\big\rangle \equiv T^{\alpha}{}_{\mu\nu}$. 
We decompose $T^{\alpha}{}_{\mu\nu}$ as~\cite{krt08} 
\begin{equation}\label{T_Decomp}
T_{\alpha\mu\nu}=\textstyle{\frac{1}{3}}(g_{\alpha\mu}T_\nu-g_{\alpha\nu}T_\mu)
-\epsilon_{\mu\nu\alpha\beta}A^\beta
+M_{\alpha\mu\nu}\,.
\end{equation}
The zero components of both $T_\mu\equiv g^{\alpha\beta}T_{\alpha\beta\mu}$ 
and $A^{\mu}\equiv\textstyle{\frac{1}{6}}\epsilon^{\alpha\beta\gamma\mu}T_{\alpha\beta\gamma}$ 
determine rotationally invariant pieces of the torsion tensor $T^{\alpha}{}_{\mu\nu}$. Our experiment is insensitive to the mixed-symmetry irreducible contribution given by
$M_{\alpha\mu\nu}\equiv\textstyle{\frac{1}{3}}(T_{\alpha\mu\nu}+T_{\mu\alpha\nu}+T_\mu g_{\alpha\nu})-\textstyle{\frac{1}{3}}(\mu\leftrightarrow\nu)$ as it is entirely anisotropic, 
but we nevertheless keep this this contribution in our theoretical analysis
for completeness.

To specify the interactions of the neutron with the background torsion pieces $T^\mu$, $A^\mu$, and $M^{\alpha\mu\nu}$ we employ a systematic approximation of possible torsion couplings to obtain the following leading-order neutron effective Lagrangian ${\cal L}_n$ of the form~\cite{krt08}:
\begin{eqnarray}\label{T_couplings}
{\cal L}_n & = & \textstyle{\frac{1}{2}}i\,\overline{\psi}\,\gamma^\mu\!\!\!\stackrel{\,\leftrightarrow}{\partial}_{\hspace{-0.7mm}\mu}\hspace{-0.9mm}\psi
-m\overline{\psi}\psi
%\nonumber\\
%&&{}
+\big[\xi_1^{(4)} T_\mu+\xi_3^{(4)} A_\mu\big]\overline{\psi}\gamma^\mu\psi
\nonumber\\
&&{}
+\big[\xi_2^{(4)} T_\mu+\xi_4^{(4)} A_\mu\big]\overline{\psi}\gamma_5\gamma^\mu\psi
%\nonumber\\
%&&{}
+\textstyle{\frac{1}{2}}i\big[\xi_1^{(5)} T^\mu+\xi_3^{(5)} A^\mu\big]\overline{\psi}\!\!\stackrel{\,\leftrightarrow}{\partial}_{\hspace{-0.7mm}\mu}\hspace{-0.9mm}\psi
\nonumber\\
&&{}
+\textstyle{\frac{1}{2}}\big[\xi_2^{(5)} T^\mu+\xi_4^{(5)} A^\mu\big]\overline{\psi}\gamma_5\!\!\stackrel{\,\leftrightarrow}{\partial}_{\hspace{-0.7mm}\mu}\hspace{-0.9mm}\psi
%\nonumber\\
%&&{}
+\textstyle{\frac{1}{2}}i\big[\xi_6^{(5)} T_\mu+\xi_7^{(5)} A_\mu\big]\overline{\psi}\sigma^{\mu\nu}\!\!\stackrel{\,\leftrightarrow}{\partial}_{\hspace{-0.7mm}\nu}\hspace{-0.9mm}\psi
\nonumber\\
&&{}
+\textstyle{\frac{1}{2}}i\epsilon^{\kappa\lambda\mu\nu}\big[\xi_8^{(5)} T_\kappa+\xi_9^{(5)} A_\kappa\big]\overline{\psi}\sigma_{\lambda\mu}\!\!\stackrel{\,\leftrightarrow}{\partial}_{\hspace{-0.7mm}\nu}\hspace{-0.9mm}\psi
%\nonumber\\
%&&{}
+\textstyle{\frac{1}{2}}i\xi^{(5)}_5M^{\nu\lambda\mu}\overline{\psi}\sigma_{\lambda\mu}\!\!\stackrel{\,\leftrightarrow}{\partial}_{\hspace{-0.7mm}\nu}\hspace{-0.9mm}\psi\,,
\end{eqnarray}
where $\psi$ denotes a Dirac spinor describing the neutron, $m$ is the neutron mass, and the $\xi^{(d)}_j$ are model-dependent couplings. The usual case of a minimally coupled point particle commonly considered in the theoretical torsion literature is recovered for $\xi^{(4)}_4 = 3/4$ with all other $\xi^{(d)}_j$ vanishing.

The mathematical structure of ${\cal L}_n$ is identical to Lagrangians employed in the study of Lorentz- and CPT-symmetry violation~\cite{sme}, as has been noted above~\cite{fn1}. 
Identifying the spacetime-constant 
$T^\mu$, $A^\mu$, and $M^{\alpha\mu\nu}$ torsion components in Eq.~(\ref{T_couplings}) 
with the background directions in Ref.~\cite{sme},
most of the mathematical machinery developed for Lorentz-violating Lagrangians 
can also be applied to ${\cal L}_n$. 
Lagrangian terms of the general structure 
$a^\mu\,\overline{\psi}\gamma_\mu\psi$, 
$\textstyle{\frac{1}{2}}ie^\mu\,\overline{\psi}\hspace{-1.7mm}\stackrel{\,\leftrightarrow}{\partial}_{\hspace{-0.7mm}\mu}\hspace{-1.7mm}\psi$,
$-\textstyle{\frac{1}{2}}f^\mu\,\overline{\psi}\gamma_5\hspace{-2.0mm}\stackrel{\,\leftrightarrow}{\partial}_{\hspace{-0.7mm}\mu}\hspace{-1.7mm}\psi$, and
$\textstyle{\frac{1}{3}}ig_\mu\,\overline{\psi}\sigma^{\mu\nu}\hspace{-2.0mm}\stackrel{\,\leftrightarrow}{\partial}_{\hspace{-0.7mm}\nu}\hspace{-1.7mm}\psi$, 
where $a^\mu$, $e^\mu$, $f^\mu$, and $g^\mu$ are spacetime constant 4-vectors, 
are known to be subdominant for a single fermion species:
first-order effects can be removed from the Lagrangian by field redefinitions~\cite{sme,redefs}. 
Comparison of these $a^\mu$, $e^\mu$, $f^\mu$, and $g^\mu$ terms 
with our Lagrangian~(\ref{T_couplings}) then establishes that 
only $\xi^{(4)}_2$, $\xi^{(4)}_4$, $\xi^{(5)}_8$, $\xi^{(5)}_9$, and $\xi^{(5)}_5$ 
can cause leading-order effects in the present context,
and we may drop all other $\xi^{(d)}_j$.

Since the measurement described below involves the motion of slow neutrons, 
the nonrelativistic limit of the physics contained in Lagrangian~(\ref{T_couplings}) is sufficient for our purposes.
To determine this limit,
we perform a generalized Foldy--Wouthuysen transformation~\cite{nonrel_limit}, 
which decouples the neutron and antineutron wave functions contained in $\psi$
and yields
\begin{equation}\label{H}
H = \frac{\vec{p}^{\,2}}{2m}+\delta \vec{b}\cdot\vec{\sigma}
\end{equation}
for the general structure  of the nonrelativistic neutron Hamiltonian.
Here, 
$\delta \vec{b}$ is determined by the background torsion
and is in general momentum dependent.
The Pauli matrices are denoted by $\vec{\sigma}$, as usual.
To present an explicit and transparent expression for $\delta \vec{b}$,
we note that 
the torsion components appear in $\delta \vec{b}$ 
in the following four combinations:
\begin{eqnarray}\label{abrreviations}
\zeta^\mu & \equiv & 
\big[2m\xi^{(5)}_8\!-\xi^{(4)}_2\big]T^\mu
+\big[2m\xi^{(5)}_9\!-\xi^{(4)}_4\big]A^\mu\,,\nonumber\\
M^j & \equiv &
m\, \xi^{(5)}_5\epsilon_{jkl}M_{0kl}\,,\nonumber\\
M_+{}^j(\hat{p}) & \equiv &
m\,\xi^{(5)}_5(M_{k0j}+M_{0jk})\,\hat{p}^k\,,\nonumber\\
M_-{}^j(\hat{p}) & \equiv &
m\,\xi^{(5)}_5\epsilon_{jkl}(M_{nkl}+2M_{0l0}\delta_{nk})\,\hat{p}^{n}\,,
\end{eqnarray}
where $\hat{p}$ is the unit momentum vector. 
With this notation, 
the background torsion $\delta \vec{b}$ takes the form
\begin{eqnarray}\label{deltab}
\delta \vec{b} & = & 
{}+\big[\vec{M}-\vec{\zeta}\,\big]
+\big[\zeta_0\hat{p}+\vec{M}_-\big]\frac{p}{m}\nonumber\\
&&{}
+\big[{\textstyle{\frac{1}{2}}}\,\hat{p}\!\cdot\!(\vec{M}-\vec{\zeta})\,\hat{p}
-{\textstyle{\frac{1}{2}}}(\vec{M}-\vec{\zeta})+\vec{M_+}\times\hat{p}\,\big]
\frac{p^2}{m^2}
%\nonumber\\
%&&{}
+{\cal O} \Big(\frac{p^3}{m^3}\Big).
\end{eqnarray}
Note that the leading-order contribution 
contained in the first square brackets above
couples just like a conventional magnetic field,
which can complicate the experimental detection 
of this particular torsion interaction. 

The present experiment involves liquid unpolarized $^4$He, 
which can only generate isotropic torsion effects on macroscopic scales.
This eliminates all torsion components from $\delta\vec{b}$
with the exception of $\zeta\equiv\zeta_0$. 
The leading torsion correction 
to the nonrelativistic neutron Hamiltonian
is then simply given by $(\zeta/m)\,\vec{\sigma}\cdot\vec{p}$.

\section{Experimental Constraints from Neutron Spin Rotation}

To derive our constraint on $\zeta$ we note that the $\vec{\sigma}\cdot\vec{p}$ term in the torsion-induced Hamiltonian violates parity and therefore causes a rotation of the plane of polarization 
of a transversely polarized slow-neutron beam 
as it moves through matter~\cite{Mic64}. 
This phenomenon is known as neutron optical activity 
in analogy with the well-known corresponding phenomenon 
of optical activity for light. 
The rotation angle $\phi_{PV}$ of the neutron spin about $\vec{p}$ per unit length $d\phi_{PV}/dL$ is known as the rotary power in light optics. 
An expression for the rotary power 
follows in an obvious way from the preceding analysis:
\begin{equation}\label{eqn2}
\frac{d\phi_{PV}}{dL}=2 \zeta \,.
\end{equation} 
This result is consistent with the expressions that one can derive in nonrelativistic scattering theory.

The experiment was performed at the NG-6 slow-neutron beamline at the National Institute of Standards and Technology (NIST) Center for Neutron Research~\cite{Nico05}. The energy spectrum of the neutrons was approximately a Maxwellian with a peak around $3\,$meV. Transversely polarized neutrons passed through 1~meter of liquid helium held at $4\,$K in a magnetically shielded cryogenic target. The apparatus sought for a nonzero spin rotation angle using the neutron equivalent of a crossed polarizer--analyzer pair familiar from light optics. The experiment, apparatus, and analysis of systematic errors has been described in detail elsewhere~\cite{Bas09, Sno11, Micherdzinska11, Swa10, Sno12}. The measured upper bound on the parity-odd neutron spin rotation angle per unit length in liquid $^{4}$He at a temperature of $4\,$K from this experiment was $d\phi_{PV}/dL=+1.7\pm9.1\textrm{(stat.)}\pm1.4\textrm{(sys)}\times10^{-7}\,$rad/m. We can therefore derive a limit on in-matter torsion directly from Eq.~(\ref{eqn2}). Our measurement 
\begin{equation}\label{result}
|\zeta|<9.1\times 10^{-23}\,\textrm{GeV}
\end{equation}
is to our knowledge the first experimental constraint on in-matter torsion. Since neutron spin rotation involves the real part of the coherent forward scattering amplitude the in-matter torsion interaction constrained in this experiment applies to an equal number of protons, neutrons, and electrons. 

One must understand that there are Standard-Model backgrounds that also can rotate the plane of polarization of the neutron from neutron interactions with electrons and nucleons, and in fact parity-odd neutron spin rotation has been observed in heavy nuclei~\cite{Forte80, Heckel82, Heckel84}. The parity-odd neutron--electron interaction is calculable in the Standard Model but is suppressed compared to neutron--nucleon parity violation by a factor of $(1-4 \sin^{2}\theta_{W}) \approx 0.1$. The quark--quark weak interactions which induce weak interactions between the neutron and the nucleons in $^{4}$He cannot yet be calculated in the Standard Model given our inability to deal with the strongly interacting limit of QCD. One can roughly estimate the expected size of NN weak-interaction amplitudes relative to strong-interaction amplitudes to be of order $10^{-6}$ to $10^{-7}$ for the slow-neutron energies used in this work, which are far below the electroweak scale~\cite{Sto74}. The best existing estimate of $d\phi_{PV}/dL$ in n-$^{4}$He from Standard-Model weak interactions was derived using existing measurements of nuclear parity violation in a specific model~\cite{Des98} and predicts $d\phi_{PV}/dL=-6.5\pm2.2\times 10^{-7}\,$rad/m. Our experimental upper bound is larger than this estimate of the Standard-Model background and we therefore ignore the unlikely possibility of a cancellation between this Standard-Model contribution and the term of interest from in-matter torsion considered in this work.

One could imagine analyzing other precision parity-violation measurements to place bounds on in-matter torsion. We expect that constraints on $\zeta$ involving neutrons could be derived from an analysis of existing measurements of parity violation in atoms sensitive to the nuclear anapole moment, which comes from parity violating interactions between nucleons~\cite{Zel57, Woo97}. The good agreement between the measurement of the weak charge of the $^{133}$Cs atom and the Standard-Model prediction~\cite{Bouchait:2005} could be used to place limits on torsion interactions involving electrons.

A further method one might employ to set experimental constraints for other components of in-matter torsion using polarized slow neutrons is to pass neutrons through a polarized nuclear target. In this case the aligned spins in the polarized target could act as a source for other components of the torsion field not considered in this work. Precision measurement of slow-neutron spin rotation through polarized nuclear targets have been extensively studied for many decades~\cite{Bar12}. The strong spin dependence of the neutron--nucleus scattering amplitude gives rise to a phenomenon referred to as nuclear pseudomagnetic precession~\cite{Bar65} in which the neutron polarization vector rotates about the axis of the nuclear polarization vector as it passes through the polarized medium. This phenomenon has been used to measure the spin dependence of neutron--nucleus scattering amplitudes for several nuclei~\cite{Abr72}, and the systematic effects in such experiments have been considered in detail due to their possible application in searches for time-reversal violation~\cite{Masuda05, Lam94}. Unfortunately the spin rotation effects in such an experiment due to nuclear pseudomagnetism from the strong neutron--nucleus interaction are quite large and at this point they are impossible to calculate from first principles, so the sensitivity of the bounds on possible in-matter torsion components would be far less stringent than those obtained in this work. However clever schemes have been discussed in the literature for the suppression of several types of systematic effects in experiments of this type~\cite{Luk11} and it is possible that an interesting experiment could be performed. 

It is also worth pointing out that a sensitive polarized-neutron transmission-asymmetry experiment using transversely polarized $5.9\,$MeV neutrons was carried out in a nuclear spin-aligned target of holmium~\cite{Huffman97} in order to search for possible P-even, T-odd interactions of the neutron. The result from this experiment was $A_{5}={\sigma_{P} \over \sigma_{0}}=+8.6\pm 7.7\textrm{(stat.+sys.)}\times10^{-6}$ where $A_{5}$ is the transmission asymmetry for neutrons polarized along and opposite a direction normal to both their momentum and to the alignment axis of the holmium nuclei. The question as to whether or not an aligned nuclear target might possess an internal torsion field different from an unaligned target and if so how it might manifest itself in this measurement is to our knowledge unexamined in the literature. However the motion of the neutron in this experiment is still nonrelativistic and the general approach of this analysis could in principle be applied in a straightforward manner. 

In the present framework,
bounds are stated on products of $\xi$ couplings and torsion components.
Care has to be taken when sensitivities of such torsion constraints 
are compared across physical systems. 
One of the reasons is that
the neutron is a composite particle.
The $\xi$ parameters in Eq.~(\ref{T_couplings})
are therefore {\em not} the universal torsion coupling constants
to elementary Dirac fermions;
they are rather to be interpreted as effective torsion couplings
pertaining to neutrons only.
Given a specific torsion model,
these effective neutron couplings
can in principle be determined from the fundamental torsion couplings
to elementary fermions.
The second reason concerns 
the size of the background torsion generated by the source matter:
it depends not only on the detailed properties of the source,
but again also on the specific torsion model.

\section{Conclusion}

Slow-neutron spin rotation is a sensitive technique to search for possible exotic neutron interactions that violate parity, especially over mesoscopic distances intermediate between macroscopic and atomic length scales. By analyzing an experimental upper bound on neutron spin rotation in liquid $^{4}$He~\cite{Sno11}, we derive what to our knowledge are the first experimental constraints on a combination of model-independent parameters that describe in-matter torsion in an unpolarized isotropic medium. It is difficult to improve our constraint by repeating the helium spin rotation measurement with greater accuracy due to the Standard-Model background discussed above expected from quark--quark weak interactions. Other atomic and nuclear parity-violation measurements might be analyzed to constrain in-matter torsion interactions of protons and electrons, and polarized slow-neutron transmission experiments through polarized and aligned nuclear targets could be analyzed within the framework presented in this Letter in order to constrain other possible in-matter torsion components. We encourage other researchers to conduct analyses of torsion searches within this more model-independent approach so that we can continue to turn the search for torsion into a more quantitative experimental science.   

\section*{Acknowledgments}

This work was supported by the DOE, by NSF grants PHY-1068712 and PHY-1207656, by the IU Center for Spacetime Symmetries, 
by the IU  Collaborative Research and Creative Activity Fund of the Office of the Vice President for Research, 
and by the IU Collaborative Research Grants program.


\begin{thebibliography}{00}
\bibitem{fn1}
We mention, 
however, 
a practical difference.
In previous studies~\cite{CL97,krt08}, 
boosts and rotations of the torsion probe 
relative to the background torsion 
would lead to apparent Lorentz- and CPT-violating effects 
that can be employed for torsion measurements.
In our experiment, 
Lorentz transformations 
of the neutrons (our torsion probe) 
relative to the helium (our torsion source)
are impractical. 
In particular, 
we expect no sidereal or annual effects.
\end{thebibliography}
\end{document}